\documentclass{aps2010} 

\usepackage{times}
\usepackage{rawfonts}
\usepackage{graphicx}

\title{MeqTrees and direction-dependent effects}

\author[org1]{\textbf{{\emph{\underline{O. M. Smirnov}}}}}
\author[org1]{\textbf{{\emph{A. G. de Bruyn}}}}

\address[org1]{ASTRON, P.O. Box 2, Dwingeloo, 7900AA, The Netherlands, smirnov@astron.nl, bruyn@astron.nl}

\begin{document}

\maketitleblock


\section{Introduction}

\begin{figure}
\begin{tabular}{@{}ccc@{}}
\includegraphics[width=8cm,clip=true,trim=0cm 4cm 0cm 2cm]{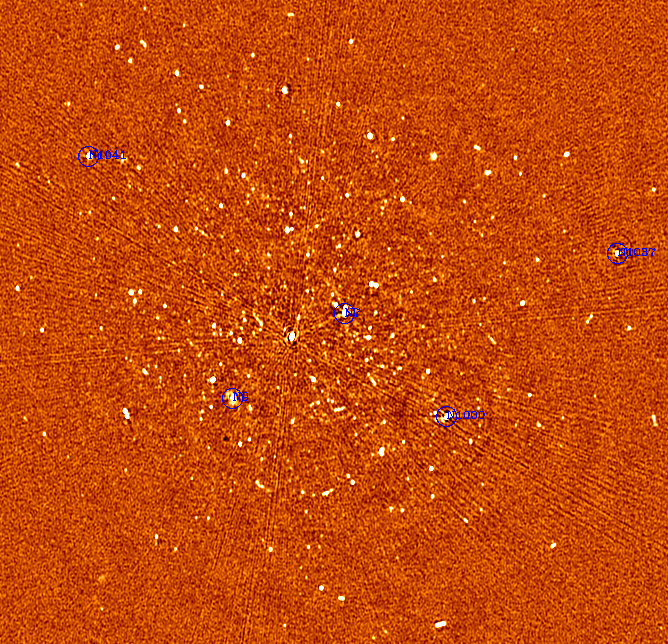} &
\includegraphics[width=8cm,clip=true,trim=0cm 4cm 0cm 2cm]{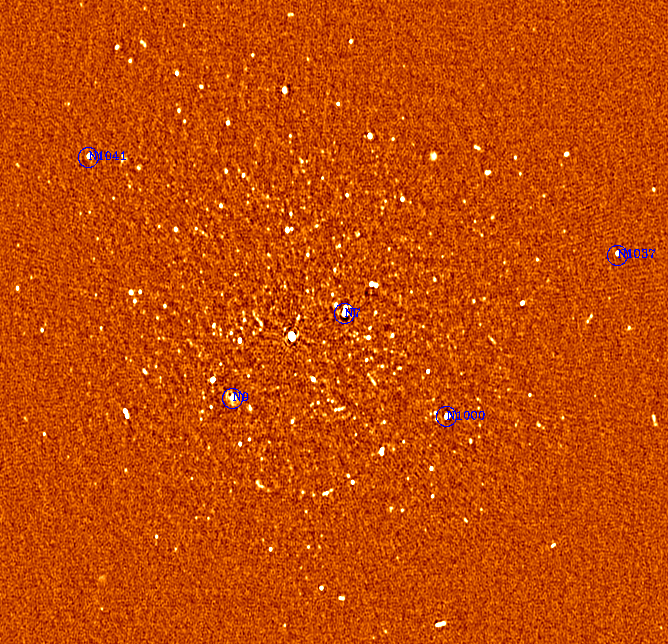} 
\end{tabular}
\caption{\label{fig:3c147}A WSRT 21cm map of 3C 147 (8-band restored image, central portion). 3C 147 itself is a 22 Jy source (below and to the left of N7), while thermal noise is 13.5 $\mu$Jy, for a dynamic range of over 1.6 million to 1. The image on the left was produced using normal selfcal in the NEWSTAR package. DDE-related artefacts around some off axis sources are clearly visible. The image on the right was produced by reprocessing the same dataset in MeqTrees, and applying differential gains to the indicated off-axis sources. (The faint radial spokes visible in the NEWSTAR map are not due to DDEs, but rather to an obscure problem with telescope RTC that was detected and eliminated during the MeqTrees reduction.)}
\end{figure}

\noindent The classical self-calibration algorithm employed in radio interferometry  associates a single direction-independent complex gain term (which can have a complicated behaviour in frequency and time) with each antenna of the interferometer. This has been spectacularly successful, and has allowed dynamic range (DR) in excess of $10^6$:1 to be reached [1], but only for relatively narrow fields with a single dominant source of emission. In many other regimes of current and future instruments, direction-dependent effects (DDEs) -- that is, effects that vary across the field of view (FoV) as well as between antennas -- become significant enough to produce DR-limiting calibration artefacts. The Westerbork Synthesis Radio Telescope (WSRT) is a particularly low-DDE instrument due to its careful design, but at sufficiently high DR, clear DDE-related artefacts are also evident  (Fig.~\ref{fig:3c147}, left). In a previous study [2], we applied the differential gains (DG) technique using the MeqTrees package [3] to eliminate these artefacts. This was very successful at producing a noise-limited image (Fig.~\ref{fig:3c147}, right), but did not explain the ultimate cause of the DDEs in the 3C 147 dataset. The per-antenna DG solutions for the 3C 147 field showed continuity in time and consistency across frequency bands [2], which was a clear indication of some ``global'' effect at work. Crucially, this also demonstrated that sources as faint as few mJy could be used to solve for differential gains, and could thus be employed as ``calibration beacons'' in the various closed-loop methods proposed for the calibration of DDEs.

Since the ultimate cause of DDEs in this observation remained unclear, we decided to conduct a separate study to narrow down the suspects. Of these, pointing error is commonly held to be the leading one (at least in WSRT 21cm observations). We therefore decided to locate a field containing a number of relatively bright off-axis point sources (the type of field radio astronomers normally try to avoid because of the calibration problems that ensue!), and observe it with deliberate pointing errors on a few of the telescopes. This was also structured as a challenge: observatory staff did not reveal to us which telescopes had been chosen for mispointing, and we had to establish this from the data itself.

\section{The QMC and QMC2 observations and reductions}

\noindent Two suitable fields were identified through an automated search of the NVSS catalogue, and designated as QMC and QMC2\footnote{In honour of the long-defunct WSRT Quality Monitoring Committee.}. Of these, QMC looked initially more promising due to the presence of a relatively strong source (1.4 Jy, compared to a maximum of 220 mJy apparent in the QMC2 field), so it was observed first. Our reduction strategy was to use the NEWSTAR package to iteratively build up a sky model during normal selfcal, then feed that model into MeqTrees, repeat selfcal, and apply DG solutions.

The QMC results proved to be only mildly satisfactory. We did obtain meaningful DG solutions, from which four mispointed antennas were correctly identified, but we could not build up a satisfactory sky model. A few of the QMC sources turned out to be slightly extended (which had not been apparent at the resolution of the NVSS), and this source structure proved to be very difficult (if not impossible) to disentangle from the severe DDEs, given the one-dimensional instantaneous PSF of the WSRT. Without an adequate sky model, the resulting images were plagued by residual artefacts. We therefore decided to concentrate on the QMC2 field instead. Having learned from the QMC experience, we requested two separate observations: one error-free 12-hour synthesis from which we intended to build a good sky model, and a second 12-hour run with deliberate mispointings.

\begin{figure}
\begin{tabular}{@{}cc@{}}
\includegraphics[width=8cm]{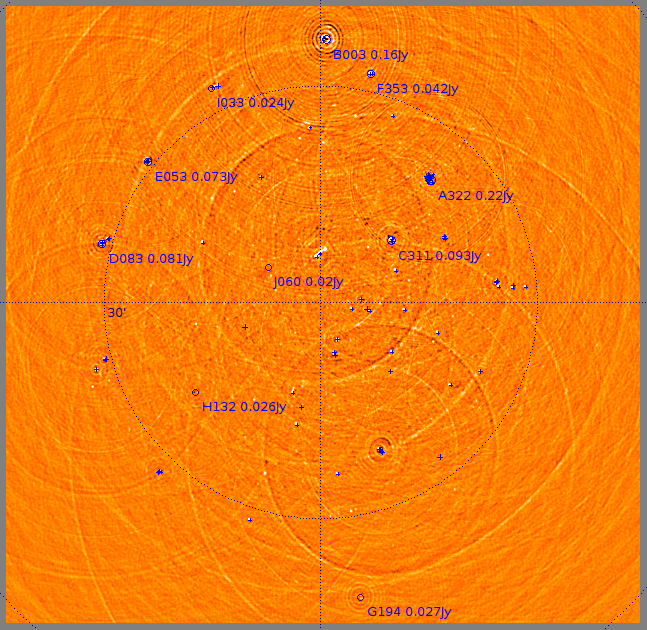} &
\includegraphics[width=8cm]{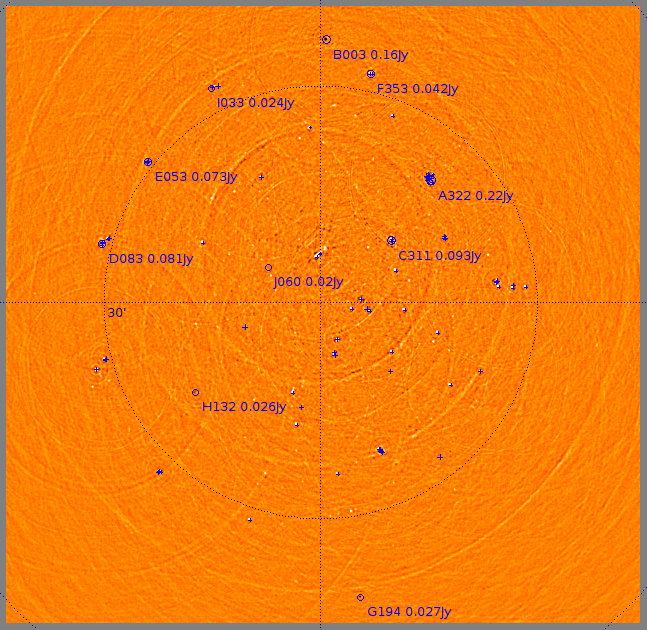} \end{tabular}
\caption{\label{fig:qmc2}Single-band residual dirty images of the QMC2 field. The image on the left is produced by regular selfcal. In the image on the right, differential gain solutions have been applied. The ten brightest sources are labelled -- note how these have been subtracted without a trace in the image on the right. The displayed intensity range is $\pm0.5$ mJy.}
\end{figure}

\begin{figure}
\begin{tabular}{@{}cc@{}}
\includegraphics[height=6.5cm]{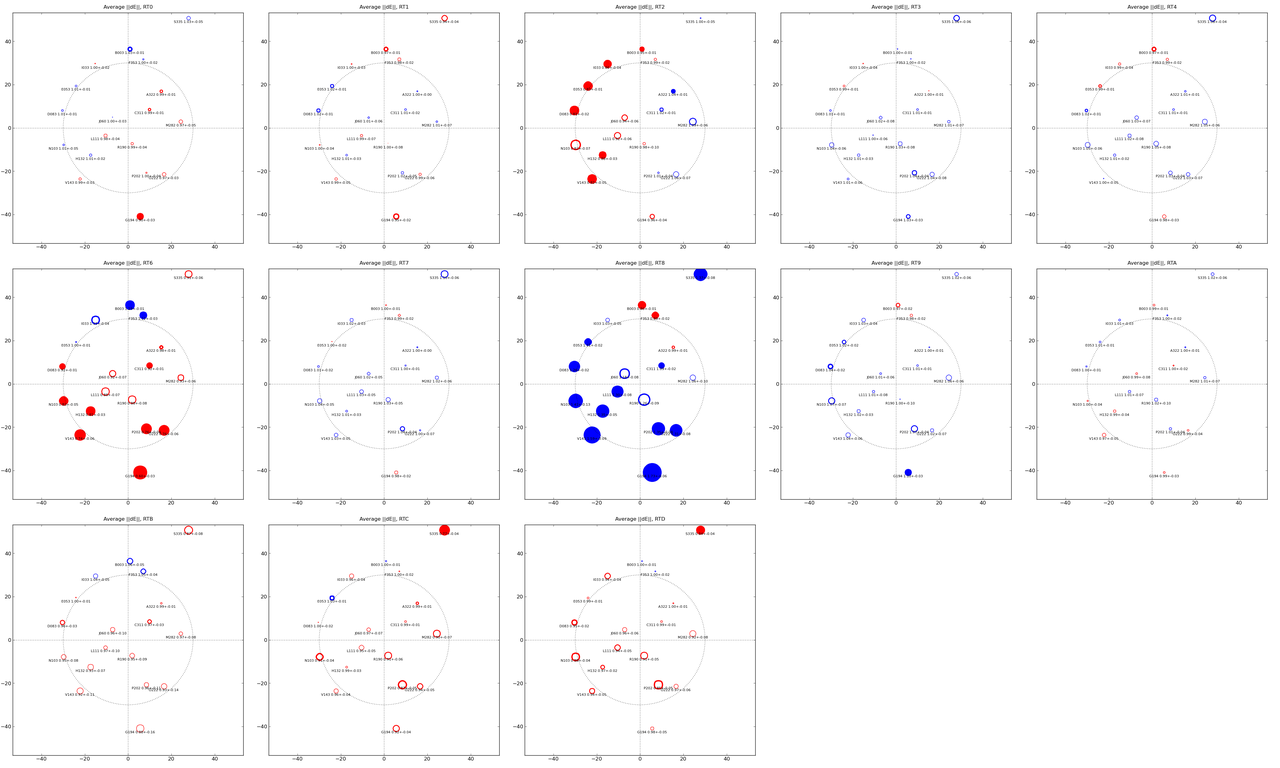} &
\includegraphics[height=6.5cm,clip=true,trim=0cm 1cm 0cm 2cm]{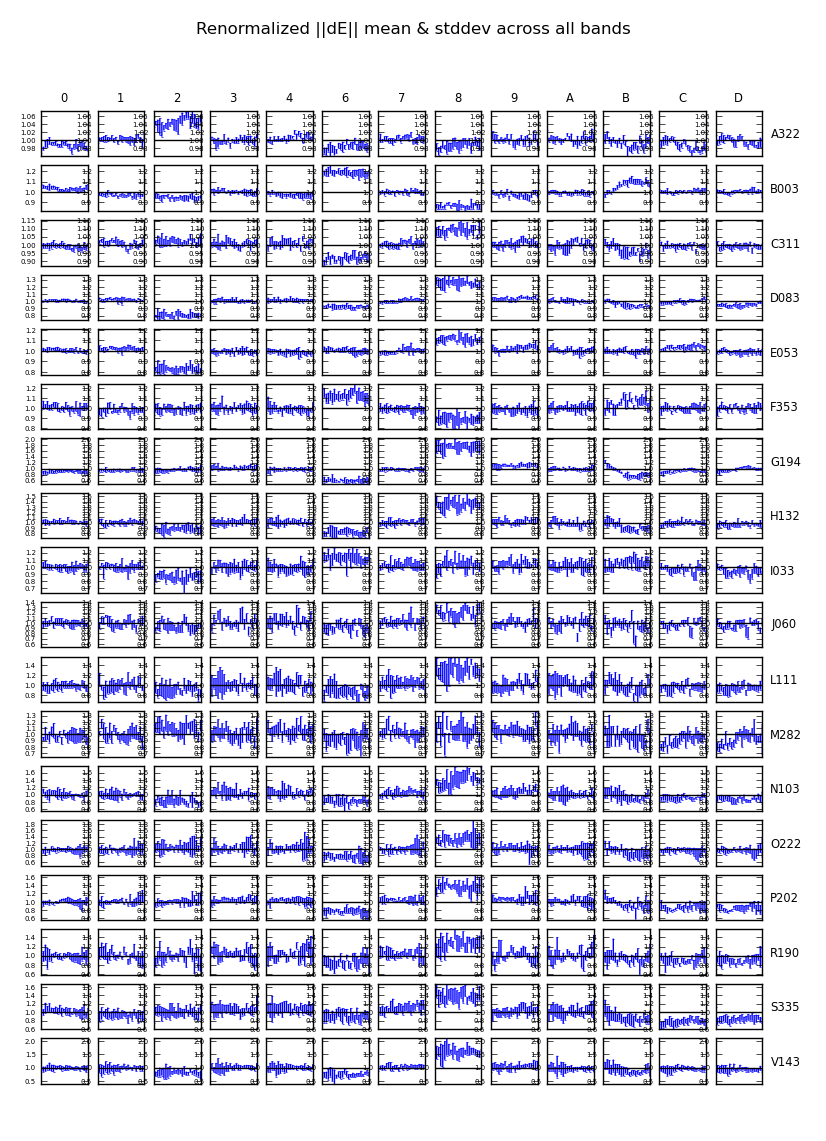} 
\end{tabular}
\caption{\label{fig:rogues}Differential gain-amplitude solutions for the QMC2 field (deliberately mispointed run). The ``Rogues Gallery'' plot on the left shows the 12-hour average DG-amplitudes. Each subplot corresponds to the field as seen by one antenna. The size of the circles indicates the deviation of each DG-amplitude above (blue) and below (red) unity. A static mispointing should, to first order, cause sources in the direction of the mispointing to appear consistently brighter (blue), and those on the opposite side fainter (red) -- this pattern can be clearly seen for antennas 2, 6 and 8. The small plots on the right show the DG-amplitudes as a function of time, per each source (across) and antenna (down). Consistent offsets in the columns corresponding to antennas 2, 6 and 8 are quite apparent. More subtle is the time-variable effect on antenna B (which is not visibile in the ``Rogues Gallery'' plot due to time averaging).
}
\end{figure}

Initial results were surprising: the error-free run looked like nothing of the sort (Fig.~\ref{fig:qmc2}, left)! DG solutions improved matters immensely (Fig.~\ref{fig:qmc2}, right), while a look at their structure suggested a serious mispointing of telescope RT8. This was communicated to the observatory staff, and they subsequently identified a malfunctioning declination encoder on the antenna, which had gone unnoticed until our experiment! Omitting baselines to RT8 allowed us to build up a good sky model of about 90 point sources. This model was then used to reduce the data from the second, deliberately mispointed run (by which point RT8 had been fixed). The resulting DG solutions (or rather their amplitudes) are shown in Fig.~\ref{fig:rogues}. From the time-averaged per-antenna plots (left), it was clear that telescopes RT2, RT6 and RT8 were mispointed towards South-West, North and South-East, respectively. More subtly, from the plots of DG-amplitude as a function of time (right), it could be determined that RTB had a time-variable mispointing. Observatory staff later confirmed these ``guesses'' were indeed correct, including the time variability of RTB (we had not asked for time variability: they had put it in on their own initiative as an extra challenge). Another subtle effect seen in Fig.~\ref{fig:rogues} is the behaviour of the northernmost source. This is actually a 3C source that is attenuated by a factor of $\sim10^3$, being almost $1^\circ$ away from the field centre. Its DG-amplitude exhibits behaviour that is opposite to the rest of the field. This is probably due to this source actually sitting on a sidelobe of the primary beam, where the gradient of the beam gain is the opposite of that on the main lobe.

As a further experiment, we tried to use MeqTrees to solve for pointing parameters directly, using the standard WSRT $\cos^3r$ model for the main lobe of the primary beam. This is conceptually similar to the ``pointing selfcal'' algorithm [4], but done via DFTs rather than FFTs. The advantage of this approach is that only two real parameters per antenna are solved for (rather than a complex gain per each source and antenna); the disadvantage is that DDEs {\em not} due to pointing error, as well as model errors, are not accounted for at all (which they are, when DGs are used). The results of this are still inconclusive at time of writing: the recovered pointing parameters are consistent with what we now know to have been the actual mispointings for the QMC2 observation, but the resulting images still suffer from significant residual artefacts (compared to those produced by the DG approach). Whether this is due to other DDEs, or to errors in our sky model, is yet to be established, but further work is ongoing.

\section{Conclusions}

\noindent Given a good sky model, the DG approach allows us to eliminate DDE-related artefacts and produce noise-limited maps even in the presence of rather severe DDEs. However, this method does not scale as well as the more global parametrizations such as pointing selfcal, so it can only be applied to a relatively small number of troublesome bright sources. On the other hand, DG provide an extremely valuable way of \emph{measuring}\/ the unknown DDEs, and thus should be used as an exploration tool in the development of new calibration methods.

\begin{figure}
\center \includegraphics[width=.9\textwidth]{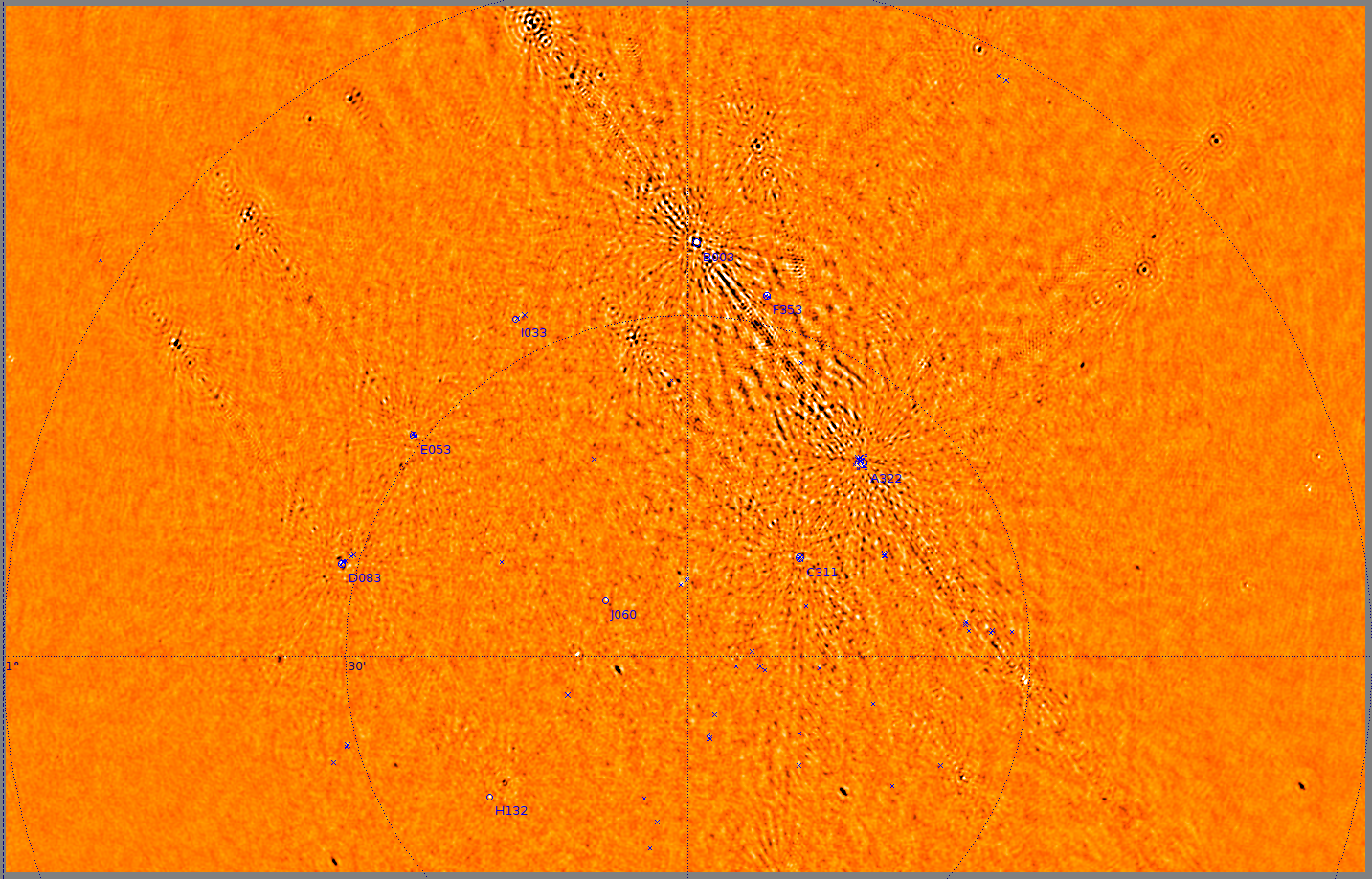} 
\caption{\label{fig:ghosts}Selfcal ``ghosts''. This shows the difference between an 8-band QMC2 map produced via selfcal and DGs, and the same map after a second round of selfcal with DG solutions in effect. Since thermal noise is the same, the difference shows calibration artefacts eliminated by the second selfcal run. The most prominent structure is the string of ``ghosts'' between the two brightest sources in the field (A322 and B003), but many ghosts associated with other sources are also evident. The displayed intensity range is $\pm75 \mu$Jy.}
\end{figure}

Finally, a very intriguing development can be seen in Fig.~\ref{fig:ghosts}. This shows an \emph{incremental difference\/} image of QMC2 after a second round of selfcal (with DGs applied). The ghost-like structures are, in essence, calibration artefacts produced by selfcal in the presence of DDEs. Such ghosts must be present in \emph{all} observations (!), but are normally masked by the thermal noise. This effect should be of great concern to currently discussed methods for the statistical detection of sub-noise signals (in particular, the Epoch of Reionization signature), and warrants further study urgently.

\section{References}

\noindent 1. A. G. de Bruyn, ``Full polarization wide field WSRT calibration/imaging at a million to one dynamic range'', 
\emph{SKA Calibration and Imaging Workshop (CALIM 2006)}, Cape Town, December 2006

\noindent 2. O. M. Smirnov, ``Revisiting the radio interferometer measurement equation. III. Addressing direction-dependent effects in 21cm WSRT observations of 3C 147'',
\emph{Astronomy \& Astrophysics}, 527, March 2011, p. A108

\noindent 3. J. E. Noordam and O. M. Smirnov, ``The MeqTrees software system and its use for third-generation calibration of radio interferometers'',
\emph{Astronomy \& Astrophysics}, 524, December 2010, p. A61

\noindent 4. S. Bhatnagar, T. J. Cornwell and K. Golap, ``Solving for the antenna based pointing errors'', EVLA Memo \#84, National Radio Astronomy Observatory, 2004, http://www.aoc.nrao.edu/evla/geninfo/memoseries/evlamemo84.pdf

\newpage

\end{document}